\documentclass[prb,superscriptaddress,floatfix,twocolumn,amsmath]{revtex4-2}
\usepackage{graphicx}
\usepackage{wasysym}
\usepackage{hyperref}
\usepackage{color}
\def\newr{\color{black}}

\def\lnod{La$_2$NiO$_{4+\delta}$}
\def\lsno{La$_{2-x}$Sr$_x$NiO$_4$}

\begin{document}

\title{Nature and impact of stripe freezing in La$_{1.67}$Sr$_{0.33}$NiO$_4$}

\author{A. M. Merritt}
\author{D. Reznik}
\affiliation{Department of Physics, University of Colorado at Boulder, Boulder, CO 80309, USA}
\author{V. O. Garlea}
\affiliation{Neutron Scattering Division, Oak Ridge National Laboratory, Oak Ridge, Tennessee 37831, USA}
\author{G. D. Gu}
\author{J. M. Tranquada}
\email{jtran@bnl.gov}
\affiliation{Condensed Matter Physics \&\ Materials Science Department, Brookhaven National Laboratory, Upton, New York 11973-5000, USA}
\date{\today}
\begin{abstract}
La$_{1.67}$Sr$_{0.33}$NiO$_4$ develops charge and spin stripe orders at temperatures of roughly 200~K, with modulation wave vectors that are temperature independent.  Various probes of spin and charge response have provided independent evidence for some sort of change below $\sim50$~K.  In combination with a new set of neutron scattering measurements, we propose a unified interpretation of all of these observations in terms of a freezing of Ni-centered charges stripes, together with a glassy ordering of the spin stripes that shows up in neutron scattering as a slight rotation of the average spin direction.
\end{abstract}
\maketitle

\section{Introduction}

The occurrence of spin and charge stripe order in \lsno\ (LSNO) is well established \cite{ulbr12b,tran13a}.  The maximum ordering temperatures occur for $x=1/3$ \cite{cheo94,rami96}, where the spin and charge orders each have a commensurate period of three lattice spacings \cite{lee97,yosh00,du00}.  Within the stripe-ordered phase, the resistivity grows quite large on cooling and a charge-excitation gap of 0.4 eV is observed to develop in optical conductivity \cite{kats96}.  Given the insulating character, one might expect that most properties would evolve monotonically with cooling; however, the system turns out to be much more interesting than that.

\begin{figure}[b]
\centerline{\includegraphics[width=0.8\columnwidth]{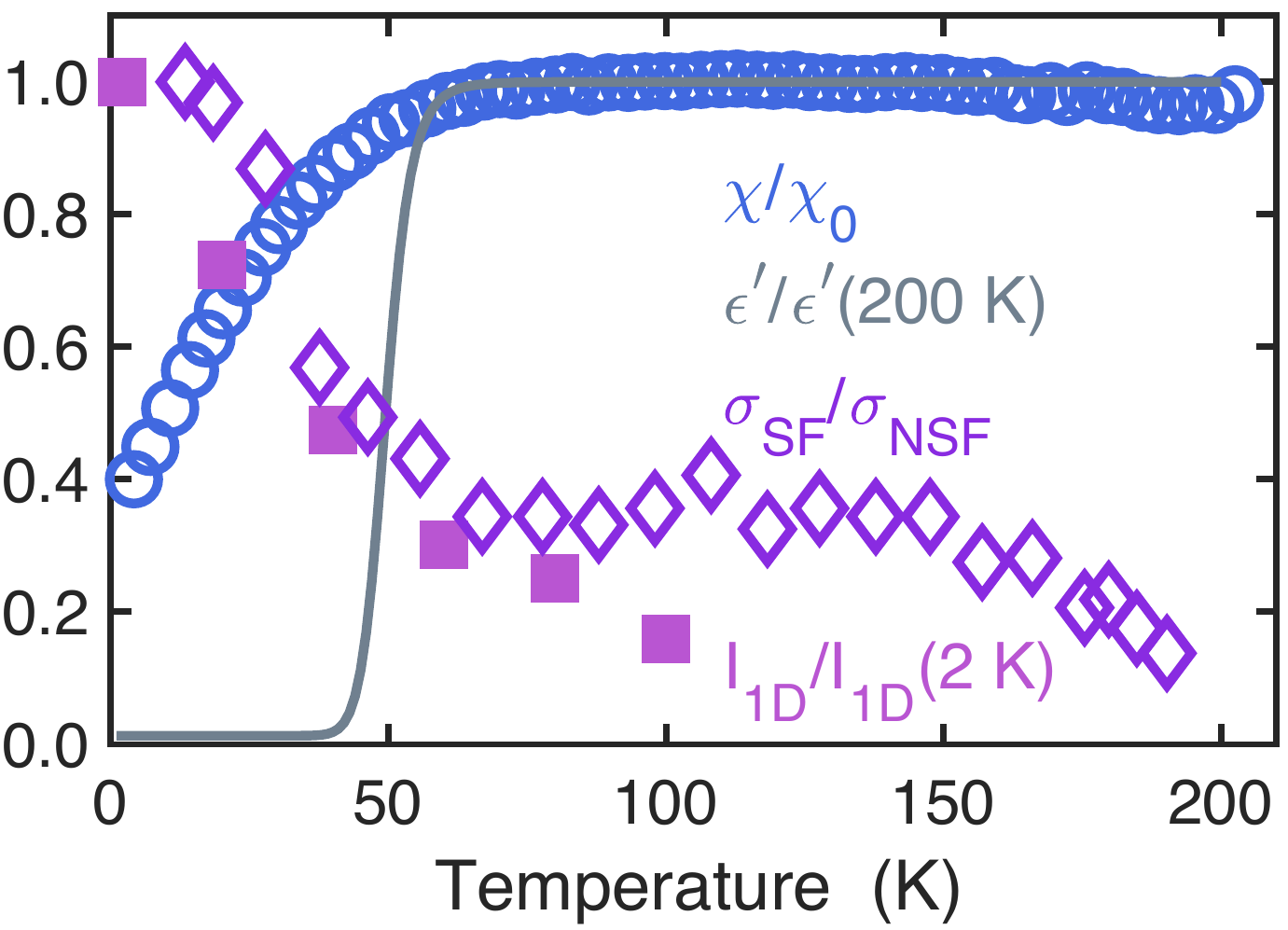}}
\caption{Temperature dependence of various measured quantities for LSNO $x=0.33$: blue circles, magnetic susceptibility $\chi(T)$, measured with an in-plane field of 1 T \cite{klin05}, divided by $\chi_0=C/(T+18.6~K)$ with $C=0.28$~emu mol$^{-1}$ K$^{-1}$; gray line, fit to the real part of the dielectric function, $\epsilon'$, measured at a frequency of 10 kHz with the electric field parallel to the planes \cite{park05}; purple diamonds, spin-flip neutron cross section relative to the non-spin-flip response measured with neutron polarization analysis at ${\bf Q}=(\frac23,0,1)$, normalized at the base temperature \cite{lee01}; violet squares, peak intensity of the 1D magnetic scattering measured for excitations of 2 meV, normalized at 2 K \cite{boot03b}. }
\label{fg:prev} 
\end{figure}

In 2003, Boothroyd and coworkers \cite{boot03b} reported the discovery of one-dimensional (1D) spin excitations within the stripe-ordered phase of LSNO $x=0.33$.  While spin stripe order sets in at 190~K \cite{lee97,klin05}, and 2D spin dynamics have been observed at temperatures well above that in closely-related samples \cite{lee02,bour03}, these 1D excitations only become significant below $\sim50$~K, as indicated by the squares plotted in Fig.~\ref{fg:prev}.  As if this were not enough, several other changes in magnetic correlations were reported to occur at similar temperatures. For example, there is a ferrimagnetic (or weak ferromagnetic) behavior in the bulk magnetic susceptibility, measured with an in-plane field, that decreases from its Curie-Weiss behavior below 50~K or so \cite{klin05}.  Also, a decrease in intensity of a small-wave-vector magnetic peak was observed below 40~K by neutron scattering \cite{lee97}.  A later study with neutron polarization analysis indicated an enhancement of the spin-flip cross section at low temperature, which was modeled in terms of a rotation in the spin direction \cite{lee01}.

\begin{figure*}[t]
\centerline{\includegraphics[width=1.9\columnwidth]{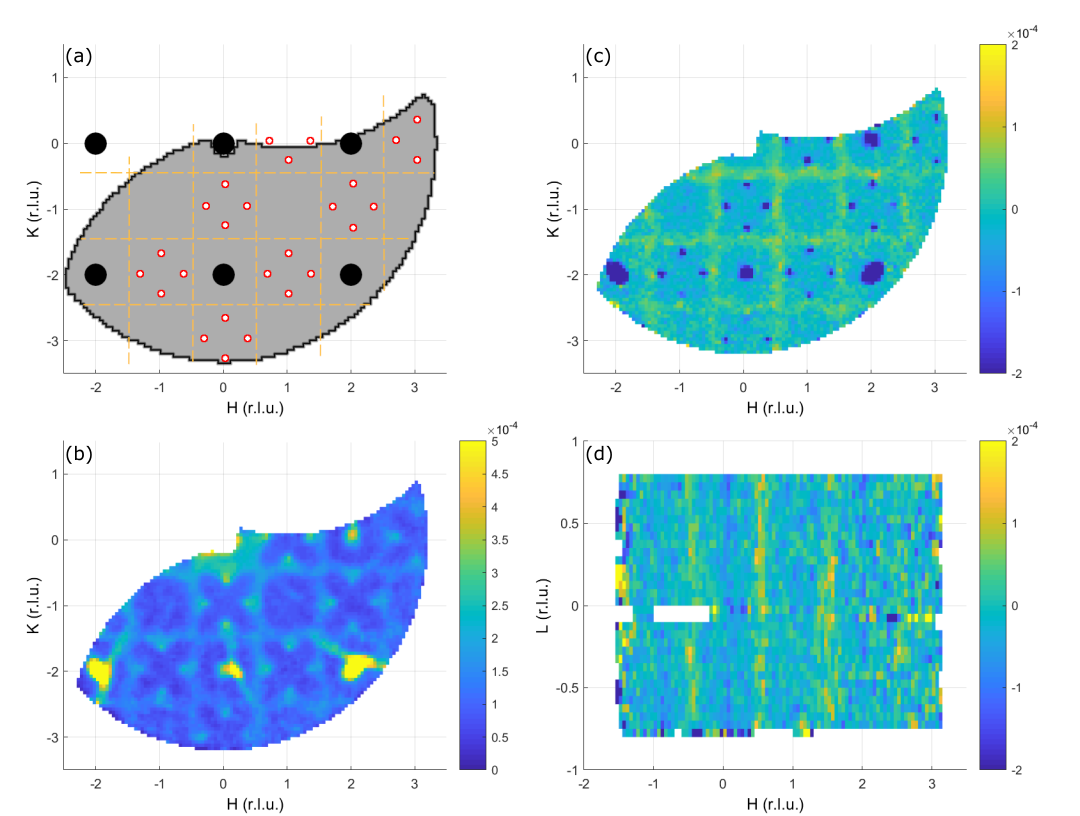}}
\caption{(a) Schematic diagram of features in reciprocal space: black circles, nuclear Bragg peaks; red circles, overlapping spin and charge stripe Bragg peaks; orange dashed lines, intersection of the sheets of scattering from individual spin stripes with the $L=0$ plane; gray shading indicates the region probed experimentally.  (b) Inelastic scattering at $\hbar\omega=3\pm1$~meV and $T=5$~K.  (c) Difference between the scattering measured at 5 K and 70 K for $\hbar\omega=3\pm1$~meV.  (d) Similar to (c), but plotting in the $(H,-0.8,L)$ plane, providing evidence that there are 2D sheets of scattering.}
\label{fg:one} 
\end{figure*}

\begin{figure}[t]
\centerline{\includegraphics[width=0.8\columnwidth]{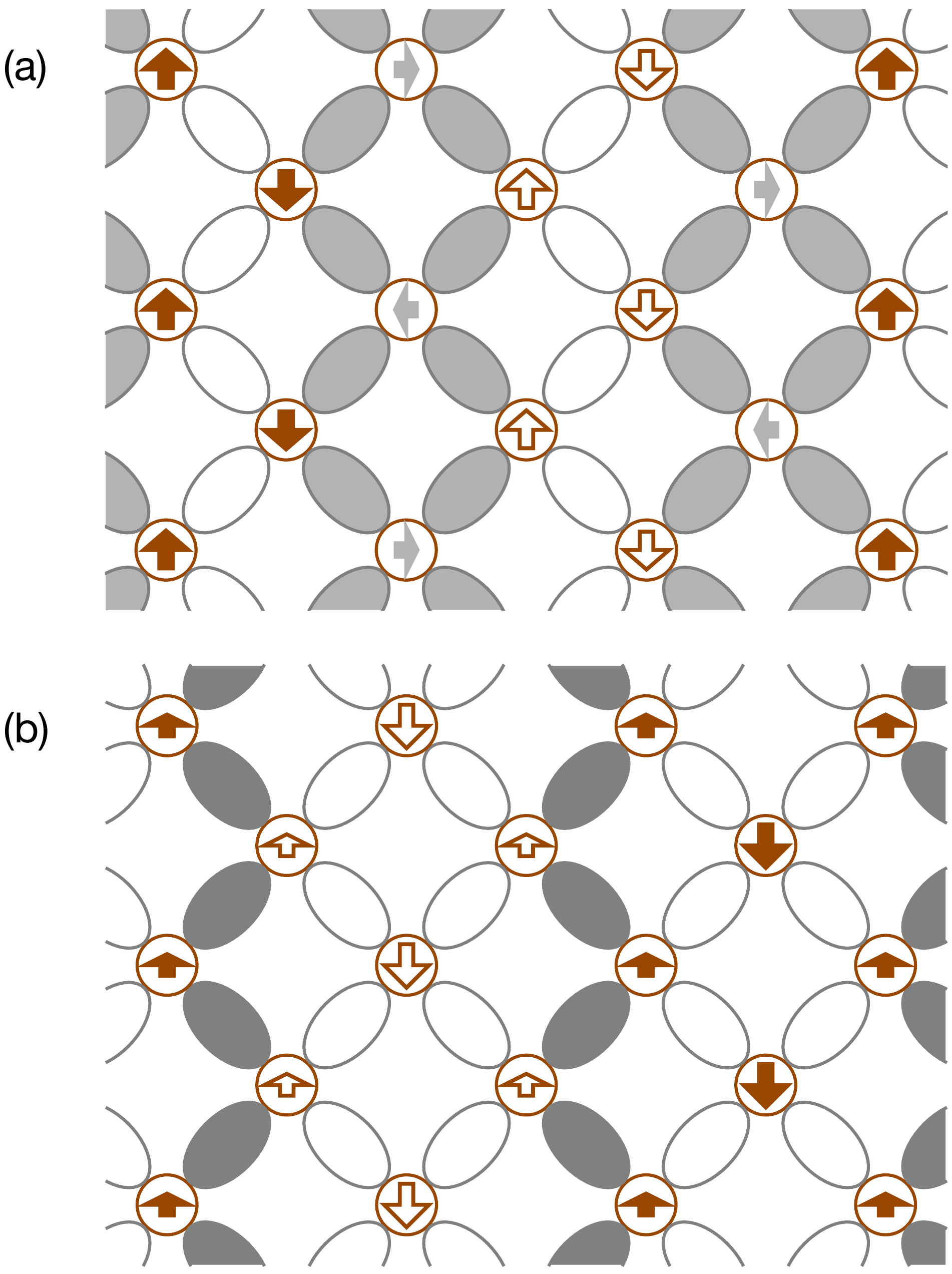}}
\caption{Schematic diagrams of charge and spin order for (a) Ni-centered stripes and (b) O-centered stripes.  Circles indicate Ni sites and arrows indicate ordered moments; ellipses correspond to O sites and gray shading indicates location of doped holes.  {\newr The intensity of the gray shading reflects the number of O sites over which a single hole is shared: 4 in (a) and 2 in (b).} Red filled/open arrows indicate antiphase spin stripes.  Gray arrows in (a) indicate relative spin correlations along the charge stripes; these correlations are purely dynamic \cite{boot03b}.  Height of arrows indicates effective spin size, as discussed in the text. }
\label{fg:versions} 
\end{figure}

We have confirmed the 1D character of the spin excitations that appear at low temperature, as shown in Fig.~\ref{fg:one} and discussed in detail below.
To make sense of the {\newr temperature-dependent} changes, we must take account of the two types of charge stripes that can occur for $x=1/3$. {\newr From previous work \cite{tran97b,woch98}, we know that charge stripes tend to be centered on columns of either Ni or O atoms, and at $x=1/3$ neighboring stripes must, on average, be of the same type. These two possibilities} are illustrated in Fig.~\ref{fg:versions}, where we associate the doped holes {\newr (gray shading)} with O sites {\newr (ellipses)}, consistent with spectroscopic studies \cite{kuip95,pell96,schu05}.  Within {\newr the} Ni-centered charge stripes {\newr of Fig.~\ref{fg:versions}(a)}, there is one hole present {\newr (shared by 4 O sites)} to hybridize with each $S=1$ Ni site, which can result in a net $S=\frac12$ {\newr (gray arrows)} \cite{schu05}.  Furthermore, the ordered Ni spins {\newr (red arrows)}  surrounding a {\newr Ni-centered} charge stripe have a staggered arrangement, so that the coupling of $S=\frac12$ moments {\newr within the charge} stripe {\newr to} neighboring $S=1$ spins is geometrically frustrated; the coupling of spins along the stripe presumably occurs via O-O hopping.  Hence, Ni-centered stripes appear to be compatible with 1D spin correlations along the charge stripes.

In contrast, the configuration with O-centered stripes{\newr, shown in Fig.~\ref{fg:versions}(b),} seems to be incompatible with 1D antiferromagnetic (AF) correlations.  In that case, the Ni spins neighboring a charge stripe are all co-aligned, and they have the same alignment for all stripes within a correlated region.  Furthermore, in order to have no average ordered moment, the spin of Ni sites adjacent to a charge stripe must be half of that of the Ni sites in the middle of a spin domain {\newr (as indicated by the reduced arrow height)}.  Of course, applying a magnetic field can change this balance, resulting in the ferrimagnetic response detected in susceptibility measurements \cite{klin05}.  (In fact, starting with a paramagnetic phase, an applied field can induce striped order with this modulation in a related nickelate \cite{tran97b}.)  

The temperature dependence of various properties must involve {\newr a} balance between the two types of stripes and the role of fluctuations.  To gain insight, we analyze the temperature dependence of about three dozen magnetic peak intensities, covering a temperature range of 5 to 70 K.  From this, we determine that we have a mixture of the two types of stripes, with a preference for Ni-centered stripes.  Also, our results are consistent with a broad distribution of spin orientations, as indicated by a nuclear magnetic resonance (NMR) study \cite{yosh99}, but this distribution is sensitive to temperature.  

To make sense of the temperature-dependent changes in the spin correlations, we have to take into account evidence for stripe freezing from other techniques.  In particular, measurements of the dynamic polarizability indicate a frequency-dependent freezing that occurs near 50 K for measurements at 10 kHz \cite{park05,fili09}.  Evidence for spin freezing at a similar temperature comes from NMR measurements on the analog compound La$_2$NiO$_{4.17}$ \cite{abus99}.  We propose that the freezing of both charge and spin degrees of freedom is necessary to realize the 1D spin correlations in the Ni-centered charge stripes, and that the freezing is reflected in the changes in the distribution of spin orientations that we observe.

The rest of the paper is organized as follows.  The experimental methods are described in the next section, followed by a presentation of the results and analysis in Sec.~III.  A discussion of the implications of the results is given in Sec.~IV, followed by our conclusions.  The appendix contains some formulas used in the analysis.

\section{Experimental Methods}
Neutron scattering measurements were performed on the time-of-flight Hybrid Spectrometer (HYSPEC) at BL-14B at the Spallation Neutron Source, Oak Ridge National Laboratory \cite{hyspec15}.  The cylindrical single crystal (4 cm in length, 0.6 cm in diameter) of LSNO $x=0.33$ was grown at Brookhaven by the travelling-solvent floating-zone method and was the subject of a previous neutron-scattering study \cite{anis14}. The sample was mounted in a Displex closed-cycle cryostat in an orientation such that wave vectors ${\bf Q}=(H,K,0)$ were in the horizontal scattering plane, with the $c$-axis vertical and perpendicular to the incident beam. To probe inelastic scattering, an incident energy $E_i$ of 27~meV was selected together with a chopper frequency of 420~Hz, resulting in an energy resolution $\Delta E\sim0.4$~meV for energy transfer $\hbar\omega=0$.  The detector vessel was positioned at $33^\circ$ to cover horizontal scattering angles from $3^\circ$ to $63^\circ$.  To map excitations in a volume of reciprocal space, the sample was rotated over a range of $150^\circ$ in steps of $0.5^\circ$.    For elastic scattering studies, we used $E_i=50$~meV and a chopper frequency of 300~Hz, with corresponding  energy resolution $\Delta E\sim1.25$~meV for $\hbar\omega=0$; this allowed us to cover an $L$ range of $\pm1.1$ reciprocal lattice units.  Data were collected for 2 detector positions, $33^\circ$ and $34.5^\circ$, with sample rotations over a $60^\circ$ range in steps of $0.5^\circ$.  (The 2 detector positions are used to compensate for detector gaps.)  Data analysis was done with the \texttt{DAVE} \cite{dave09}, \texttt{Mantid} \cite{mantid14} and \texttt{HORACE} \cite{horace16} software packages.

LSNO has a tetragonal crystal structure with space group $I4/mmm$, but the analysis is more straightforward when using a doubled unit cell volume corresponding to space group $F4/mmm$; we will do so here. In this case, $a=b=5.42$~\AA\ and $c=12.7$~\AA, and we will use reciprocal lattice units (rlu) $(2\pi/a,2\pi/a,2\pi/c)$.  The scattering features of interest are illustrated schematically in Fig.~\ref{fg:one}(a); these will be discussed in more detail in the next section. 

Determination of the relative intensities of the elastic magnetic peaks was done by integrating the total intensity within a cylindrical volume of radius 0.15 rlu in the $(H,K)$ plane about each spin-order peak, and extending to $\pm0.25$ rlu in $L$. To determine the background, signal was integrated in  contiguous volumes adjacent in the transverse in-plane direction; note that the background tends to vary significantly with $\vert \textbf{Q} \vert$ but much less with $\hat{\bf Q}={\bf Q}/Q$. This background was subtracted from the total integrated intensity to obtain the integrated intensity of each peak.

\section{Results and Analysis}

Based on previous studies of related nickelate samples \cite{tran95b,woch98,huck06}, we expect that the stripe order is locally unidirectional; however, because the average crystal structure is tetragonal, we average over stripe twin domains.  For one domain, peaks due to charge and spin order are allowed at ${\bf Q} = {\bf G}+{\bf g}_{\rm co}$ and ${\bf G}+{\bf g}_{\rm so}$ for nuclear Bragg peaks {\bf G}, with ${\bf g}_{\rm co}=(2\epsilon,0,L)$ and ${\bf g}_{\rm so}=(1\pm\epsilon,0,L)$.  For LSNO $x=0.33$, it happens that $\epsilon=0.33$, so that the charge and spin peaks overlap \cite{lee97,yosh00,boot03b,anis14}.  The magnetic scattering is strongest at small $Q$, while the charge-order signal only becomes significant at larger $Q$ \cite{anis14,zhon17}.  In the following, we focus on relatively small $Q$ and ignore charge-order contributions, treating them as a systematic error.

We start by showing that the 1D excitations are distinct from the 3D magnetic order that originates from the spin stripes; instead, they are consistent with antiferromagnetic correlations among moments residing in charge stripes centered on Ni-sites, in which each Ni moment is reduced by a low-spin hybridization with a doped hole.  We then focus on the thermal evolution of the 3D magnetic order.

Within the $(H,K,0)$ plane, the inelastic 1D scattering shows up as lines running along $H$ and centered at $K=m+0.5$ for integer $m$; the response from the twin domain is rotated by $90^\circ$, as indicated in Fig.~\ref{fg:one}(a).  An actual measurement at 5~K, integrating excitations between 2 and 4 meV, is shown in Fig.~\ref{fg:one}(b).  The strongest scattering comes from acoustic phonons about the nuclear Bragg peaks; there are also spurious arcs of scattering near these peaks due to Bragg scattered neutrons that have been scattered a second time, possibly from the front detector window.  Next strongest are the spin excitations dispersing from the spin-order wave vectors \cite{boot03a,woo05}.  The weakest excitations are the lines of 1D scattering, that appear to form a square grid.  On warming, the 1D excitations lose intensity, while the spin-stripe excitations, corresponding to normal modes of the spin-stripe order, grow in intensity.  Hence, by subtracting a measurement at 70 K from the 5-K data, we obtain Fig.~\ref{fg:one}(c).  Here, the 1D signal is positive, while the signal about the spin-order peaks is negative (similar to the acoustic phonons).  To confirm that the 1D scattering is truly one-dimensional, Fig.~\ref{fg:one}(d) shows that the inelastic signal at $H=n+0.5$ ($n$ integer) is independent of $L$.

To understand the modulation wave vector of the 1D excitations, consider the schematic diagram of Ni-centered stripes in Fig.~\ref{fg:versions}(a), focusing on the Ni sites (open circles) within a charge stripe.  If there is a net $S=1/2$ on each Ni site, with antiferromagnetic correlations between nearest neighbors, the period is $2a$ along the charge stripe, corresponding to $K=0.5$ in reciprocal space.  The associated spin excitations disperse along this direction, but not in the transverse directions.  We have confirmed the dispersion originally determined by Boothroyd {\it et al.} \cite{boot03b}, but as we did not improve on their results, we show no data.

\begin{figure}[t]
\centerline{\includegraphics[width=1.\columnwidth]{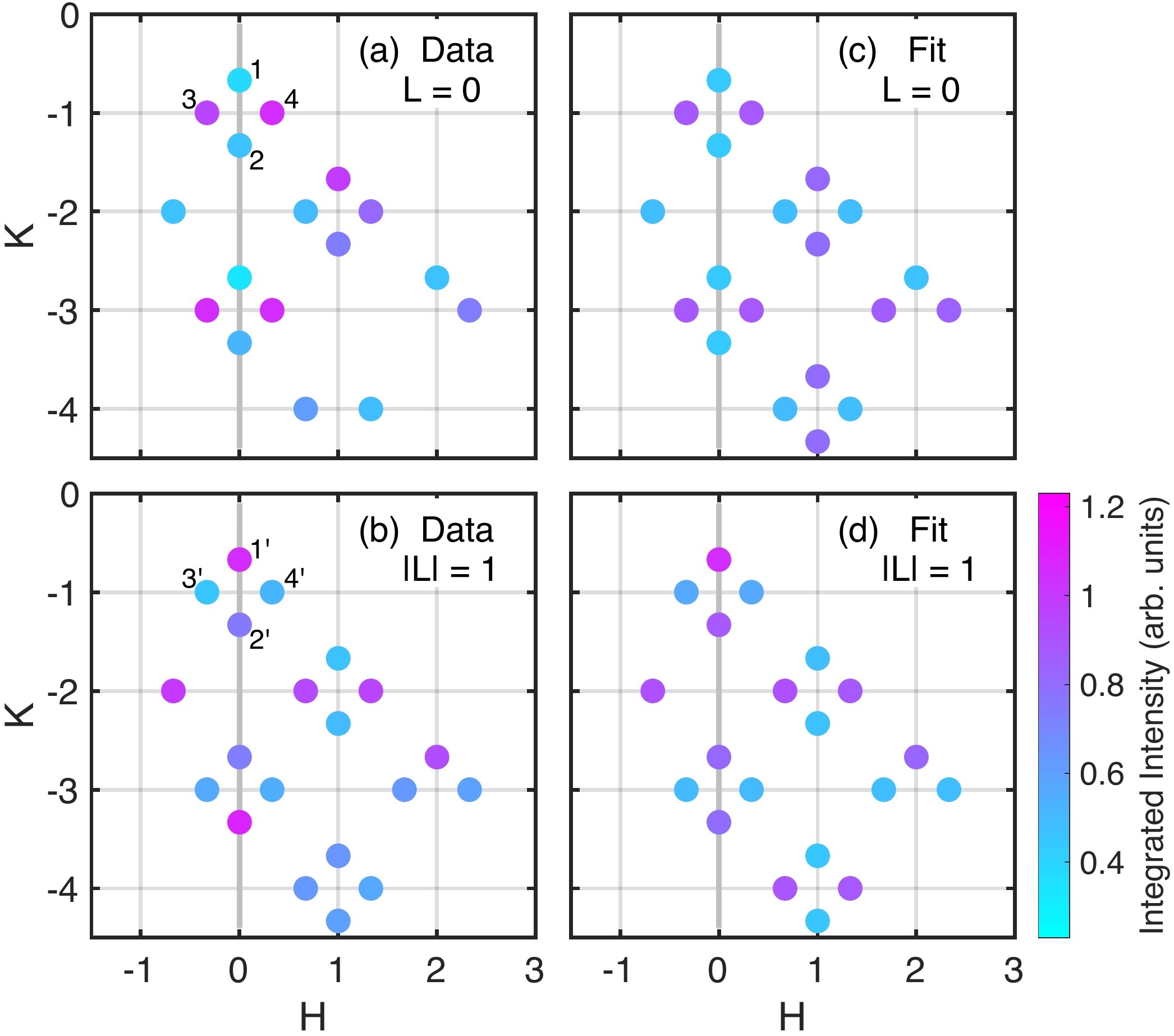}}
\caption{Color map of measured magnetic peak intensities at $T=5$~K for (a) $L = 0$ and (b) an average of $L=\pm1$; also included are fitted intensities for (c) $L=0$ and (d) $|L|=1$.  For (a), 3 measured points with anomalously large intensities, possibly from charge-order contributions, have been excluded.}
\label{fg:mag} 
\end{figure}

To examine the temperature-dependent changes of the 3D magnetic peak intensities, we consider the integrated intensities for elastic scattering at a number of 3D magnetic peak positions.  (The scattering is actually broad in $L$ \cite{lee97} due to stripe stacking disorder \cite{tran96a}, but we simply integrated over $L_0\pm0.25$ where $L_0 =-1$, 0, or $+1$.)  Examples of peak intensities measured at 5~K are shown in Fig.~\ref{fg:mag}(a) and (b).  

The scattered intensity is proportional to:
\begin{equation}
 {\cal S}({\bf Q}) = |F({\bf Q})|^2 |f({\bf Q})|^2 \langle \sin^2 \eta \rangle,
\end{equation}
where $F({\bf Q})$ is the structure factor, $f({\bf Q})$ is the magnetic form factor, $\eta$ is the angle between ${\bf Q}$ and the spin ${\bf S}$, and $\langle\ldots\rangle$ emphasizes that this quantity is averaged over the sample.  The magnetic form factor for Ni in La$_2$NiO$_4$ was found to be remarkably constant for $Q \lesssim 3.8~\mbox{\AA}^{-1}$ \cite{wang92} (corresponding to 3.3 rlu within the NiO$_2$ planes); for simplicity, we will assume it to be constant over our measurement range.  In terms of spin direction, we know that the spins tend to align perpendicular to the stripe modulation direction when there is long-range order, as in the case of La$_2$NiO$_{4.133}$ \cite{tran95b,woch98}.  We expect that the spins are locally collinear [and assume this in calculating $F({\bf Q})$] but that the spin axis may vary locally due to various defects \cite{yosh99}.  We will return to the distribution of spin orientations below.

Let us assume for the moment that the sample has a single type of stripe order, which we label N for Ni-centered charge stripes or O for O-centered.  Then for stripe order $j$ ($j=$ N, O), we find
\begin{equation}
  F({\bf Q}) = F_j({\bf Q})\left[1+e^{i\pi(H+L)}\right],
  \label{eq:F1}
\end{equation}
where $F_j({\bf Q})$ is the structure factor for a single layer of order $j$, and the term in brackets is determined by the stacking from one layer to the next.   Note that Coulomb repulsion favors a centered stacking, which would involve a shift of $0.75a$ from one layer to the next; with a single type of stripe, the best that can be achieved is a shift of $0.5a$.  The stacking factor in Eq.~(\ref{eq:F1}) corresponds to the latter case.

With O-centered stripes, as illustrated in Fig.~\ref{fg:versions}(b), there are two inequivalent Ni sites within each spin stripe.  The central Ni site is farthest from the doped holes, and we will assume that it has $S=1$.  The two Ni sites adjacent to the charge stripes are likely to have reduced effective spins.  In order to avoid a net ferromagnetic moment, these sites must then have $\langle S\rangle=1/2$.  As there is only half a doped hole per edge Ni, this reduction cannot be achieved by low-spin hybridization alone; fluctuations of the spins due to local fluctuations of the charge stripe may also play a role.  For Ni-centered stripes, there are two equivalent Ni sites, as shown in Fig.~\ref{fg:versions}(a); in this case, we assume that each hole is already compensating a Ni spin within the charge stripe.  For the later analysis, however, it is convenient to have $|F_{\rm N}({\bf Q})| = |F_{\rm O}({\bf Q})|$ for all allowed ${\bf Q}$, in which case we reduce the net spin for $F_{\rm N}$ to $\sqrt{3}/2$.  Formulas for $F_{\rm N}({\bf Q})$ and $F_{\rm O}({\bf Q})$ are given in the Appendix.

Before we consider general fits to the measured intensities, we first need to discuss the distribution of spin orientations.  For this purpose, it is useful to start with a simple analysis of a set of low $Q$ peaks labeled 1--4 and $1'$--$4'$ in Fig.~\ref{fg:mag}(a) and (b), respectively.  From previous work \cite{lee97}, we know that the peaks at ${\bf G}+{\bf g}_{\rm so}$ are strong for $L=1$ and weak for $L=0$, related to the character of the stripe stacking.  To take account of the fact that we see peaks from twinned stripe domains, it is convenient to alter the notation of ${\bf g}_{\rm so}$ to ${\bf g}_{L}=(\frac23,0,L)$ and ${\bf g}_{L}'=(0,\frac23,L)$, with $L=0$ or 1.  In Table~\ref{tab:Q}, we identify the magnetic modulation wave vectors associated with the labeled peaks ${\bf Q}_j$.  From an examination of the table, one can see that we should get the small  structure factor magnitude for points ${\bf Q}_1$, ${\bf Q}_2$, and ${\bf Q}_{3'}$, ${\bf Q}_{4'}$, and the large structure factor magnitude for ${\bf Q}_3$, ${\bf Q}_4$, and ${\bf Q}_{1'}$, ${\bf Q}_{2'}$.  Comparison with the color-coded intensities in Figs.~\ref{fg:mag}(a) and (b) shows approximate consistency with this; however, we have not yet taken account of the factor $\langle \sin^2 \eta\rangle$, which should vary among otherwise-equivalent points.

\begin{table}[b]
\caption{\label{tab:Q} Decomposition of the wave vectors ${\bf Q}_i$ noted in Fig.~\ref{fg:mag}(a) and (b) into ${\bf G}+{\bf g}_{\rm so}$, where ${\bf G}$ is a reciprocal lattice vector and ${\bf g}_{\rm so}$ is a spin-order modulation.  As discussed in the text, the possible modulation wave vectors are ${\bf g}_0 = (\frac23,0,0)$ and ${\bf g}_1=(\frac23,0,1)$ for one stripe domain, and ${\bf g}_0'=(0,\frac23,0)$ and ${\bf g}_1'=(0,\frac23,1)$ for the other.}
\begin{ruledtabular}
\begin{tabular}{clcrc}
 & ${\bf Q}$ & ${\bf G}$  & ${\bf g}_{\rm so}$ & \\
\hline
 & ${\bf Q}_1$ & $(0,0,0)$ & ${\bf g}_0$ & \\
 & ${\bf Q}_2$ & $(2,0,0)$ & $-{\bf g}_0$ & \\
 & ${\bf Q}_3$ & $(-1,-1,-1)$ & ${\bf g}_1'$ & \\
 & ${\bf Q}_4$ & $(1,-1,1)$ & $-{\bf g}_1'$ & \\
 & ${\bf Q}_1'$ & $(0,0,0)$ & ${\bf g}_1$ & \\
 & ${\bf Q}_2'$ & $(2,0,2)$ & $-{\bf g}_1$ & \\
 & ${\bf Q}_3'$ & $(-1,-1,1)$ & ${\bf g}_0'$ & \\
 & ${\bf Q}_4'$ & $(-1,1,1)$ & $-{\bf g}_0'$ & \\
\end{tabular}
\end{ruledtabular}
\end{table}

The points ${\bf Q}_1$ and ${\bf Q}_2$ are parallel to the stripe modulation, while ${\bf Q}_{3'}$ and ${\bf Q}_{4'}$ are transverse.  Since the absolute magnitude of the structure factor is identical for all of these points, the ratio
\begin{equation}
  R_0 = {I({\bf Q}_1)+I({\bf Q}_2)\over I({\bf Q}_{3'})+I({\bf Q}_{4'})}
  \label{eq:r0def}
\end{equation}
should depend only on the values of $\langle \sin^2 \eta\rangle$ at these points.  We get another combination with a different dependence on $\langle \sin^2 \eta\rangle$ with the ratio
\begin{equation}
  R_1 = {I({\bf Q}_3)+I({\bf Q}_4)\over I({\bf Q}_{1'})+I({\bf Q}_{2'})}.
  \label{eq:r1def}
\end{equation}
We expect the spins to lie in the plane, and we can reference the orientation of spin components to the spin modulation vector ${\bf g}_{\rm so}$, which is along [100].  Assuming a normalized distribution, we take the average fraction of spin components along [100] to be $\sin^2\theta$ and that along [010] to be $\cos^2\theta$.  We then have
\begin{equation}
  \langle\sin^2 \eta\rangle = 1 - \hat{Q}_x^2\sin^2\theta - \hat{Q}_y^2\cos^2\theta,
\end{equation}
where $\hat{Q}_\alpha=Q_\alpha/Q$.  Formulas for $R_0$ and $R_1$ in terms of $\sin\theta$ are given in the Appendix.

\begin{figure}[t]
\centerline{\includegraphics[width=0.8\columnwidth]{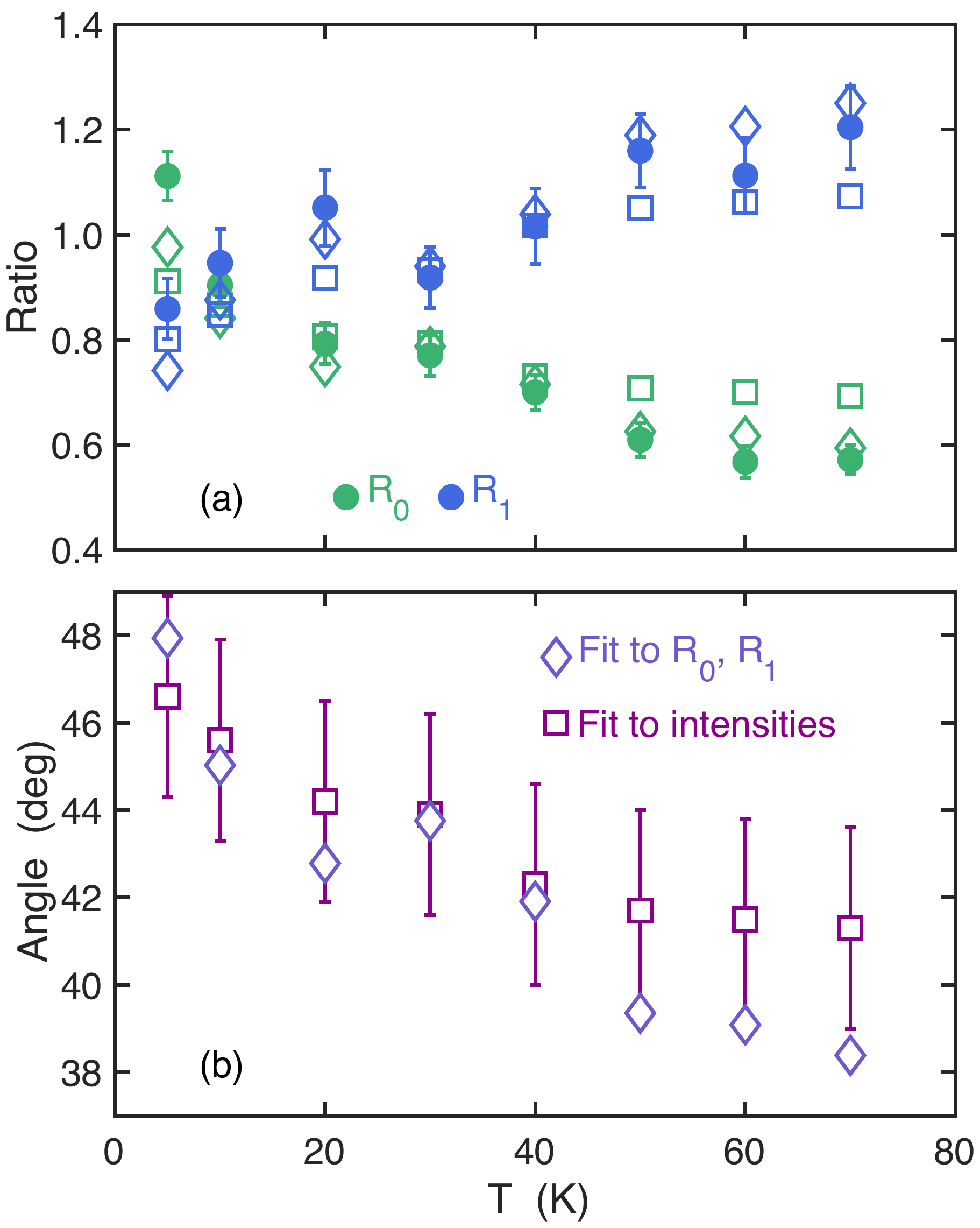}}
\caption{Top: Temperature dependence of intensity ratios $R_0$ (green) and $R_1$ (blue), defined in the text.  Filled circles: data; open diamonds: fit to $R_0$ and $R_1$ using a single $\theta$ at each temperature; open squares: results from fitting entire set of peaks at each temperature.  Bottom: Values of $\theta$ obtained by fitting just $R_0$ and $R_1$ (diamonds) and from fitting the entire data set (squares). }
\label{fg:tdep} 
\end{figure}

In Fig.~\ref{fg:tdep}(a), the experimental ratios $R_0$ and $R_1$, represented by filled circles, are plotted vs.\ temperature.  Fits to Eqs.~(\ref{eq:r0}) and (\ref{eq:r1}), indicated by open diamonds, yield the temperature dependent $\theta(T)$ shown in Fig.~\ref{fg:tdep}(b).  The magnitude of $\theta$ is always close to $45^\circ$, which means that the spin distribution has comparable spin components parallel and perpendicular to ${\bf g}_{\rm so}$.  With cooling, the distribution changes such that the average 
perpendicular spin component is growing relative to the parallel.

\begin{figure}[t]
\centerline{\includegraphics[width=0.8\columnwidth]{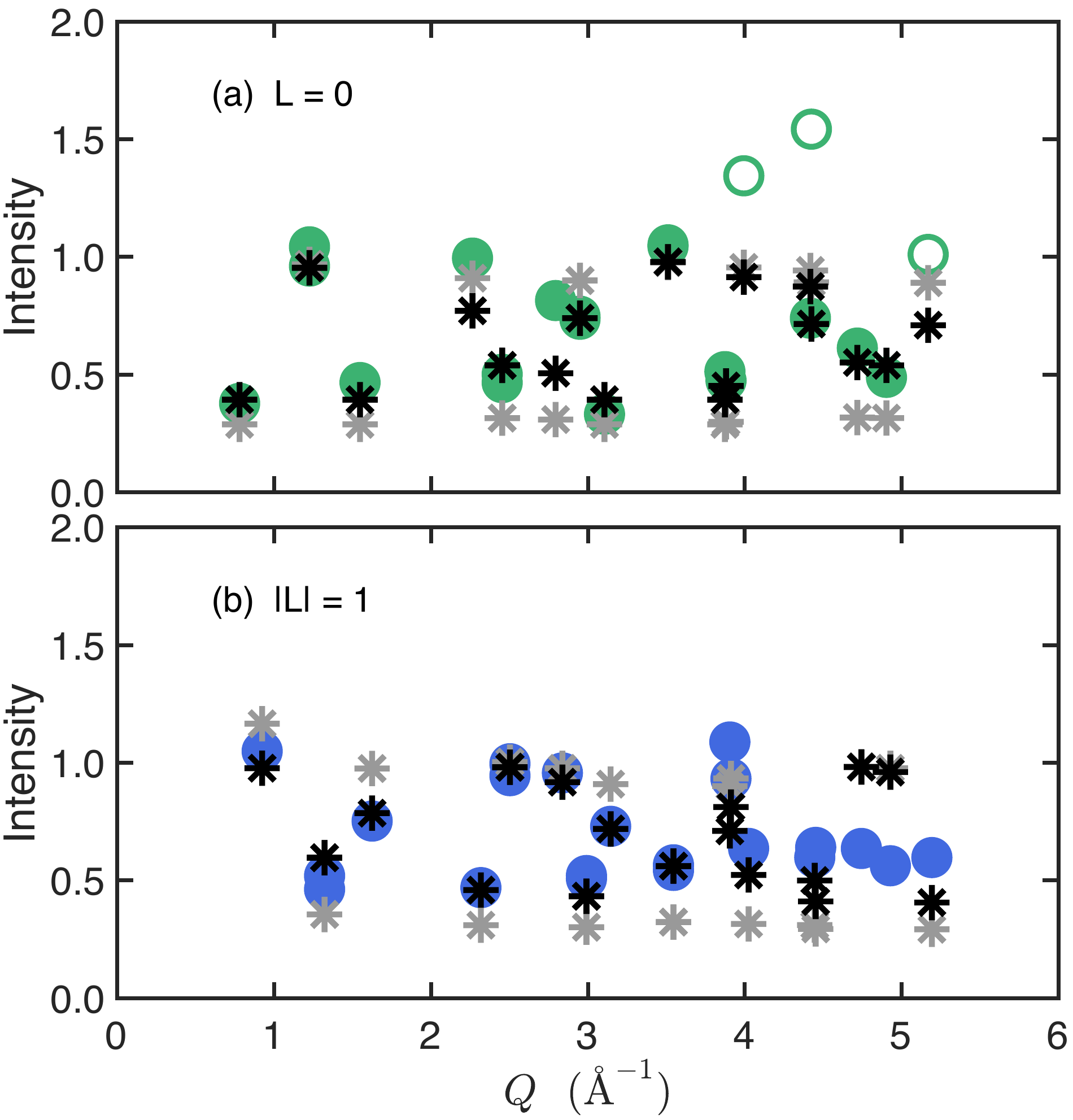}}
\caption{Fit to integrated intensities from measurements at 5~K for (a) $L=0$, and (b) average of intensities at $L=\pm1$.  Filled circles are data points used in the fit, open circles are measurements not included in fitting; gray stars represent calculated intensities for a model with $p\equiv1$, while black stars correspond to the model with variable $p$ and resulting parameter values $p=0.863\pm0.015$, $\theta = 46.6^\circ \pm2.3^\circ$. }
\label{fg:comp} 
\end{figure}

To fit all of the measured peaks, we need to take account of the structure factor and the distribution of spin orientations. We find that the quality of the fit is improved if we allow for the possible presence of both types of stripes.  Let $p$ be the probability that a given layer has Ni-centered stripes. Then we can write the structure factor as:
\begin{equation}
  F({\bf Q}) = [pF_{\rm N}({\bf Q}) + (1-p)F_{\rm O}({\bf Q})]\left[1+e^{i\pi(H+L)}\right].
  \label{eq:f}
\end{equation}
In fitting the data, we initially took $p$, $\theta$, and an overall amplitude factor as adjustable parameters.  From the fits to data at all temperatures, we found that the amplitude factor changes relatively little, so we fixed it to an average value and repeated the fits with just two parameters.  The measured (circles) and fitted intensities (black stars) for the 5-K results are shown in Fig.~\ref{fg:comp}; for comparison, a fit with $p$ constrained to be 1 is shown by gray stars.  

A qualitative comparison of the measured and fitted results is presented in Fig.~\ref{fg:mag}. The fitted results for $\theta$ are indicated by squares in Fig.~\ref{fg:tdep}(b); as one can see, they are fairly similar to the results obtained from analyzing just the first few low-$Q$ peaks.  The result for $p$ is 0.863 at 5~K, decreasing gradually to 0.847 at 70~K.   This change is small and comparable to the uncertainty of 0.015, but at least it does go in the direction of favoring Ni-centered stripes at low temperature, where the 1D spin excitations are strong, while shifting towards more O-centered stripes at higher temperatures, where the ferrimagnetic response is more robust.

\section{Discussion}

Over the years, there have been a number of theoretical calculations of stripe order and stability in LSNO \cite{zaan94,hott04,racz06,yama07,schw08,pete18}.  While they generally find diagonal stripes with one hole per Ni site to be the ground state, there has not been much discussion of temperature-dependent changes in stripe correlations or responses to external fields.

As mentioned in the introduction, many experimental studies of LSNO $x=0.33$ have focused on the evolution of either the spin or charge response.  We have argued that a proper understanding of those experiments and our own requires a simultaneous consideration of the spin and charge correlations.  The interpretation is made somewhat more complicated by glassy behavior of both the charge and spin components \cite{yosh99,park05}.

Behavior is more sharply defined in samples of \lnod, where ordered O interstitials allow long-range stripe order \cite{tran95b,woch98,abus99,abus01}.  In the case of \lnod\ with $\delta=0.133$, the low-temperature spin orientation is perpendicular to ${\bf g}_{\rm so}$ \cite{tran95b,woch98}.  Furthermore, the stripe density in that system changes with temperature, and modeling indicates that the stripes tend to be O-centered at high temperature, shifting towards Ni-centered at low temperature \cite{woch98}. When doping is achieved by randomly distributed Sr ions substituting for La, it results in finite correlation lengths for the stripe order \cite{yosh00}.  Dislocations of charge stripes, in particular, lead to local frustration of spin order and a distribution of spin orientations \cite{zaan01}.  An NMR study provides evidence for a broad distribution of spin orientations in LSNO $x=0.33$ \cite{yosh99}.

In their neutron scattering study of LSNO $x=0.33$, Lee {\it et al.} \cite{lee01} started with an assumption of a unique spin direction at higher temperatures and found evidence for a rotation (by an arbitrary amount) on cooling to low temperature.  Our analysis has similarly identified a rotation of the average spin orientation; however, that orientation is never uniquely transverse to ${\bf g}_{\rm so}$.  We have argued that it is more reasonable to assume a distribution of locally-collinear spin orientations, as indicated by NMR \cite{yosh99}, with a change in the average on cooling below 50~K.

The observation of a frequency-dependent freezing of the dielectric response \cite{park05} indicates that the charge stripes can be displaced by a dynamic electric field at higher temperatures, but that they become frozen in a glassy fashion on cooling below $\sim50$~K.  Such a change in the charge dynamics should also impact the spin correlations.  Indeed, Freeman {\it et al.} \cite{free06} studied memory effects in bulk susceptibility for a magnetic field applied parallel to the planes and found that a large irreversibility set in below $\sim50$~K, corresponding with the decrease in the ferrimagnetic response \cite{klin05}.  In our nominally-elastic neutron scattering measurements, we integrate over the low frequency ranges over which those studies detect freezing.  As a consequence, we observed a gradual change in correlations rather than a well-defined freezing transition.

The net magnetic moment associated with the ferrimagnetic (or weak ferromagnetic) response in bulk magnetization is small ($\sim0.002$~$\mu_{\rm B}$/Ni) at saturation, and disappears in zero field \cite{klin05}.  We have argued that this response likely comes from O-centered stripes; however, the modest induced moment would not require a high density of O-centered stripes, consistent with our data analysis, which indicates a density of $\sim15$\%\ for $T>50$~K.  Of course, it is possible that the applied in-plane field can induce a shift of stripes toward O-centering for temperatures above 50~K.

The shift in correlations at low temperature is towards Ni-centered stripes with the spin component perpendicular to ${\bf g}_{\rm so}$ growing.  Taking into account the glassy behavior, this is consistent with the conditions expected to support the 1D spin excitations from charge stripes.

\section{Conclusion}

Through neutron scattering measurements on a crystal of LSNO $x=0.33$, we have confirmed the development of dynamic 1D spin correlations associated with Ni-centered charge stripes.  From integrated intensities of elastic magnetic peaks associated with the 3D spin stripe order, we have obtained a new characterization of the temperature evolution of the distribution of spin orientations.  We find that a variety of distinct measurements of temperature dependent behavior reported in the literature can be interpreted consistently in terms of a combined glassy freezing of charge and spin stripes at $\sim50$~K.

\section{Acknowledgments}

AMM and DR (GDG and JMT) were supported by the U. S. Department of Energy (DOE), Office of Basic Energy Sciences, Division of Materials Sciences and Engineering, under Award DE-SC0006939 (Contract No.\ DE-SC0012704).  This research used resources at the  Spallation Neutron Source, a DOE Office of Science User Facility operated by the Oak Ridge National Laboratory

\section{Appendix}

Let the structure factor components $F_{\rm N}$ and $F_{\rm O}$ denoted in Eq.~(\ref{eq:f}) correspond to the correlations illustrated in Fig.~\ref{fg:versions}(a) and (b), respectively.  Formulas for them are as follows:
\begin{equation}
  F_{\rm N} = 2i\alpha_{\rm N}\left[e^{i\pi k}\sin(\pi h) - \sin(2\pi h)\right],
\end{equation}
where $\alpha_{\rm N} = \sqrt{3}/2$ is the assumed relative magnitude of the Ni spins adjacent to Ni-centered stripes,  and
\begin{align}
  F_{\rm O} &= 2\alpha_{\rm O} \left(e^{-i\pi h}+ e^{i\pi(2h+k)}\right)\cos(\pi h) \nonumber\\
  & \quad\quad - \left(e^{i\pi(-h+k)}+e^{i2\pi h}\right),
\end{align}
where $\alpha_{\rm O}=0.5$ is the relative magnitude of the Ni moments adjacent to an O-centered charge stripe.

The intensity ratios $R_0$ and $R_1$ are defined in Eqs.~(\ref{eq:r0def}) and (\ref{eq:r1def}), respectively.  Explicit formulas for these ratios in terms of the angle $\theta$ between the average spin direction and ${\bf g}_{\rm so}$ are as follows:
\begin{align}
  R_0 &= {1+8\sin^2\theta\over 10-5A_0\sin^2\theta}, \label{eq:r0}\\
  R_1 &= {1-\sin^2\theta\over A_1+B_1\sin^2\theta} \label{eq:r1}
\end{align}
where
\begin{align}
  A0 &= {4\over 4+9r^2}+{16\over 16+9r^2}, \\
  A1 &= {1+9r^2\over 10+9r^2}, \\
  B1 &= {8\over10+9r^2},
\end{align}
and $r=c^\ast/a^\ast$.

\bibliography{lno,theory,neutrons}

\end{document}